\documentclass[oldversion]{aa}
\usepackage{amsmath}
\usepackage{natbib}
\usepackage{footnote}

\usepackage{graphicx}

\def\url#1{{\tt#1}}

\begin{document}

\title{Cool stars in the Galactic Center as seen by  APOGEE}
\subtitle{M giants, AGB stars and supergiant stars/candidates}

\author{M. Schultheis \inst{1} 
  \and A. Rojas-Arriagada \inst{2,3}
  \and K.~Cunha \inst{4,5}  
  \and M. Zoccali \inst{2,3}
  \and C. Chiappini \inst{6,7}
  \and G. Zasowski \inst{8}
  \and A.B.A. Queiroz \inst{6,7}
  \and D. Minniti \inst{9,3,10}
  \and T. Fritz \inst{11,12}
  \and D.A.~Garc{\'i}a-Hern{\'a}ndez \inst{11,12}
  \and C. Nitschelm \inst{13}
  \and O. Zamora \inst{11,12}
  \and S. Hasselquist \inst{8,14}
  \and J.G. Fern\'andez-Trincado \inst{15}
  \and R.R. Munoz\inst{16}
}

   \institute{ Universit\'e C\^ote d'Azur, Observatoire de la C\^ote d'Azur, Laboratoire Lagrange, CNRS, Blvd de l'Observatoire, F-06304 Nice, France
 e-mail: mathias.schultheis@oca.eu
 \and
 Instituto de Astrof\'isica, Facultad de F\'isica, Pontificia, Universidad Cat\'olica de Chile, Av. Vicu\~na Mackenna 4860, Santiago,  Chile
 \and
 Millennium Institute of Astrophysics, Av. Vicu\~na Mackenna 4860, 782-0436 Macul, Santiago, Chile
\and
University of Arizona, Tucson, AZ 85719, USA
\and
Observat'orio Nacional, Sao Crist\'ovao, Rio de Janeiro, Brazil
\and
Leibniz-Institut f\"ur Astrophysik Potsdam (AIP), An der Sternwarte 16, 14482 Potsdam, Germany
\and
Laborat\'{\o}rio Interinstitucional de e-Astronomia, - LIneA, Rua Gal. Jose\'e Cristino 77, Rio de Janeiro, RJ - 20921-400, Brazil
\and
Department of Physics \& Astronomy, University of Utah, Salt Lake City, UT, 84112, USA
\and
Departamento de Ciencias Fisicas, Facultad de Ciencias Exactas, Universidad Andres Bello, Av. Fernandez Concha 700, Las Condes, Santiago, Chile
\and
Vatican Observatory, V00120 Vatican City State, Italy
\and
Instituto de Astrof\'{i}sica de Canarias, Calle Via L\'{a}ctea s/n, E-38206 La Laguna, Tenerife, Spain
\and
Universidad de La Laguna (ULL), Departamento de Astrofísica, E-30206 La Laguna, Tenerife, Spain
\and
Centro de Astronom{\'i}a (CITEVA), Universidad de Antofagasta, Avenida Angamos 601, Antofagasta 1270300, Chile
\and
$\dagger${NSF Astronomy and Astrophysics Postdoctoral Fellow}
\and
Instituto de Astronom\'ia y Ciencias Planterias, Universidad de Atacama, Copayapu 485, Copiap\'o, Chile
\and
Departamento de Astronomia, Universidad de Chile, Camino El Observatorio 1515, Las Condes, Santiago, Chile
 }


%

\abstract {The  Galactic Center region, including the nuclear disk,  has  until recently been largely avoided in  chemical census studies because of extreme extinction and stellar crowding. Large, near-IR spectroscopic surveys, such as the Apache Point Observatory Galactic Evolution Experiment  (APOGEE), allow the measurement of metallicities  in the inner region of our Galaxy. Making use of the latest APOGEE data release (DR16), we are able for the first time to study  cool AGB stars and supergiants in this region. The stellar parameters of five known AGB stars and one supergiant star (VR 5-7) show that their location is well above the tip of the RGB.
 We study  metallicities of 157 M giants situated within 150\,pc of the Galactic center from observations obtained by the APOGEE survey with reliable stellar parameters from the APOGEE/ASPCAP pipeline making use of the cool star grid down to 3200\,K. Distances, interstellar extinction values, and radial velocities  were checked to confirm that these stars are indeed situated in the Galactic Center region.
 We detect a clear bimodal structure in the metallicity distribution function, with a dominant  metal-rich peak of $\rm [Fe/H] \sim +0.3\,dex$ and a metal-poor peak around $\rm [Fe/H]= -0.5\,dex$, which is 0.2\,dex poorer than Baade's Window. The $\rm \alpha$- elements  Mg, Si, Ca, and O show a similar trend to the  Galactic Bulge. The metal-poor component is enhanced in the $\alpha$-elements, suggesting that this population could be associated with the  classical bulge and  a fast formation scenario. We find a clear signature of a rotating nuclear stellar disk and  a significant fraction of high velocity stars with $\rm v_{gal} > 300\,km/s$;  the metal-rich stars show a much higher rotation velocity ($\rm \sim 200\,km/s$) with respect to the metal-poor stars ($\rm \sim 140\,km/s$).
 The chemical abundances as well as the metallicity distribution function suggest that the nuclear stellar disc and the nuclear star cluster show distinct chemical
  signatures and might be formed differently.}



\maketitle

\titlerunning{ }
\authorrunning{ }

\section{Introduction}
 The Milky Way bulge is such a complex system that its formation and evolution are still poorly understood. As a result of high extinction and crowding, the study of the Galactic bulge remains challenging. Extinction of more than 30\,mag in $\rm A_{V}$  in the Galactic center (GC) regions requires IR spectroscopy. 
While an increasing number of detailed chemical abundances in the intermediate
and outer bulge (such as Baade's Window) are now available thanks to  large spectroscopic surveys such as ARGOS (\citealt{freeman13}), Gaia-ESO (\citealt{rojas2014}), and APOGEE (\citealt{garciaperez2013}), chemical abundances of stars in the inner Galactic bulge (IGB) with projected distances of $\rm R_{G} \leq 200\,pc$ from the GC remain poorly studied.

There have been a few earlier works dedicated to studying the chemistry of stars in the central part of the Milky Way. Using high-resolution IR spectra ($R\sim$40,000),  \citet{carr2000} made the first detailed abundance measurement of a star in the Galactic center within 1\,pc, the M2 supergiant IRS 7, finding roughly solar metallicity and an abundance pattern that was consistent with the dredge-up of CNO cycle products. \mbox{\citet{Ramirez2000}} also analyzed high-resolution IR spectra (R=40,000) of a sample of ten cool supergiant and red-giant stars in the GC and estimated an iron abundance  of $+$0.12 $\pm$ 0.22\,dex. Subsequently, \citet{cunha2007} used $R\sim$50,000 spectra to  study the detailed chemistry of the \citet{Ramirez2000} sample and found a slightly enhanced and narrow metallicity distribution ($\rm \langle [Fe/H] \rangle =+0.14 \pm 0.06\,dex$) that was also alpha-enhanced: note that this small sample of 10 stars, save for one, were members of the central cluster. More recently, Ryde \& Schultheis (2015, RS2015) studied nine field stars with projected distances of $\rm R \leq 50\,pc$ and also found a metal-rich population with  $\rm [Fe/H]= +0.11 \pm 0.15\,dex $ and a lack of a metal-poor population (similarly to Cunha et al. 2007), but found low [$\alpha$/Fe] values, except for calcium. Their mean metallicities are $\sim$ 0.3\,dex higher than fields in the inner bulge (\citealt{rich2007}; \citealt{rich2012}), indicating that the GC region contains  a distinct population. A refined analysis (\citealt{Nandakumar2018}) gave a slightly higher mean metallicity of $\rm [Fe/H]= +0.3 \pm 0.10\,dex$ and confirmed the very narrow distribution. \citet{grieco2015} compared these data with a chemical evolution model and concluded that in order to reproduce the observed $\rm [\alpha/Fe]$ ratios, the GC region should have experienced a main strong burst of star formation and should have  evolved very quickly with an IMF  which contains more massive stars.

\citet{ryde:16} presented an abundance study of 28 M giants in fields located within a few degrees south of the GC using high-resolution ($R\sim$50,000) spectra. They found a wide range of metallicities that narrows towards the center and confirmed the \citet{rich:12} study that found alpha enhancement throughout the inner bulge. This would be consistent with a homogeneous enrichment history in the inner bulge.
More recently, \citet{schultheis2019} found evidence of a chemically distinct  population in the nucleus (nuclear stellar disc and nuclear star cluster)  compared to the Inner Galactic Bulge that shows a very high fraction of metal-rich stars  ($\sim$ 80\%) with a  mean  metallicity of +0.2\,dex in the GC. 
They concluded that the nucleus mimics
a metallicity gradient of $\rm \sim -0.27\,dex/kpc$ in the inner 600\,pc but is flat if  the GC is excluded.

\citet{Schultheis2015}[hereafter referred to as S15]  studied 33 M giant stars of APOGEE in the so-called GALCEN field to study the metallicity distribution function and chemical abundances using the DR12 data release (\citealt{apogee_dr12}). 
However, they had to limit their analysis to stars with $\rm T_{eff} > 3700\,K$,  due to problems with modeling of the coolest stars  (\citealt{holtzman2015}), thus excluding a large number of cool stars- including asymptotic giant branch stars (AGB) and supergiants- from their analysis. They found some  evidence
of the presence of a metal-poor fraction of stars that are $\alpha$-enhanced.


In this paper we take advantage of the recently released and improved results from  APOGEE DR16 (\citealt{Ahumada2019}), which has several improvements over previous data releases, in particular, DR12 (see details in J\"onsson et al. 2020) and includes additional observations of one plate at APOGEE-S that is dedicated to observations of targets in the Galactic center. It is most relevant to the study of cool luminous stars in the central parts of the Milky Way that the DR16 results now extend to effective temperatures as low as 3200K using the grid of spherical MARCS model atmospheres (\citealt{MARCS}). In this paper we present results for the metallicity distribution function and chemical abundances of 157 M giants, which constitutes one of the largest high-resolution spectroscopic  samples in the inner degree of the Milky Way to date. In addition, this is the first study of cool AGB stars and supergiants in the GALCEN field using results from the APOGEE survey.


\section{The sample} \label{sec:sample}

\subsection{APOGEE}
The Apache Point Observatory Galactic Evolution Experiment (APOGEE)  is a large scale, near-IR,  high-resolution ($R \sim 22,500$) spectroscopic survey of   the Milky Way 
stellar populations, mainly focused on red giants (\citealt{Majewski2017,zasowski2013,zasowski2017}).  APOGEE has been a component of both SDSS-III and SDSS-IV \citep{eisenstein2011,Blanton2017} and uses custom-built twin spectrographs at Apache Point Observatory's 2.5~m Sloan Telescope and Las Campanas Observatory's 2.5~m du~Pont telescope \citep{gunn2006,Wilson2019}.
APOGEE observes in the $H$-band, 
where extinction by dust is significantly lower than at optical wavelengths 
(e.g., $A(H) / A(V) \sim 0.16$). 

With its high resolution and high S/N ($\rm \sim 100$ per Nyquist-sampled pixel),
APOGEE determines both accurate radial velocities (better than $\rm 0.5\,km\,s^{-1}$) and
reliable  abundance measurements, including the most abundant metals in the universe (C, N, O), along with other $\alpha$, odd-$Z$, and iron-peak elements.
The latest SDSS-IV Data Release (DR16; \citealt{Ahumada2019})  provides to the scientific community  spectra of more than 430,000 stars, as well
as the derived stellar properties, including radial velocities, effective temperatures, surface gravities, metallicities, and individual
abundances. Additional information, such as photometry and target selection criteria, is also provided and described in \citet{zasowski2013,zasowski2017}.
Stellar parameters and chemical abundances for up to 24 elements were  derived by the APOGEE Stellar Parameters and Chemical Abundances Pipeline (ASPCAP, \citealt{garcia-perez2016}), while in \citet{Nidever2015}  the data reduction and radial velocity pipelines are described.
The model grids are based on a complete set of MARCS stellar atmospheres (\citealt{MARCS}), and the spectral synthesis using the  Turbospectrum code (\citealt{Alvarez1998},\citealt{Plez2012}) with a  new updated line list has been used (Smith et al.  2020, submitted).

In addition to DR16, we use additional observations  including additional stars observed after those released in DR16 from APOGEE-S. These data were reduced with the same pipeline as the DR16 stars.

\subsection{Distances} \label{distances}


We obtained the  StarHorse (SH) distances from Queiroz et al. (\citealt{SH})  for the latest APOGEE-2 survey data release DR16, as well as for the additional stars not present in DR16.  SH is a Bayesian fitting code that derives distances and extinction but also other quantities such as ages and masses.  For more details about the method, the priors, and its validation, we refer to \citet{Queiroz2018} and to  Queiroz et al. (\citealt{SH}) which combines the ASPCAP stellar parameters with Gaia DR2 and photometric surveys achieving in this way precise distances and extinction;
the extinction treatment has been significantly improved in the updated version.
The typical internal precision in distance for  our GALCEN DR16 sample is about $\rm 10 \pm 7 \%$.  SH uses   the APOGEE targeting extinction estimate (see \citealt{zasowski2013}, \citealt{zasowski2017})  as a prior for the total line-of-sight extinction.  In addition, we also did a comparison with the 3D extinction map of \citet{schultheis2014} using VVV data and the 2D extinction map of \citet{gonzalez2012} where we find similar $\rm A_{K}$ values (within 10\% uncertainty) with respect to RJCE.

\subsection{The GALCEN field} \label{sample}

Two plates of GALCEN, one at APO (location ID 4330) and one at LCO (location ID 5534),  have been observed with a total number of 619 stars. Due to the high interstellar extinction, a  special target selection procedure has been applied for the GALCEN field (\citealt{zasowski2013}). Known  AGB long-period variables based on K-band lightcurves  (\citealt{matsunaga2009}) have been targeted as well as spectroscopically identified  supergiants such as IRS7, IRS19, and IRS22 and supergiant candidates based on photometric color-color criteria.  These AGB and supergiant  targets have among the coolest effective temperatures ($\rm T_{eff} <   3500 K$) in the APOGEE stellar sample.  While in S15 due to the limiting ATLAS9 model atmospheres grid in ASPCAP for cool stars,  only 33 warm K/M giants were used, we make use in this paper  of the cooler model grid of MARCS model atmospheres to quadruple the sample.

One of the most significant improvement of DR16 compared to DR14 is the accuracy and consistency of the derived stellar parameters for the coolest giants ($\rm T_{eff} < 3500\,K$) which is due to the use of MARCS model atmospheres in spherical symmetry. This improvement can be clearly seen by the visual inspection of the HR diagram (see e.g. Fig.~\ref{stelparams}) , especially the ``clumpiness'' of the HR diagram with $\rm T_{eff} < 3500\,K$ seen in DR14 is now gone (see \citealt{Joensson2020}).  However, ``external'' calibrators  for these cool stars are lacking which would be important   to estimate  the accuracy of the stellar parameters for those objects.

For our selection, we disregarded stars with ASPCAPFLAGS=='STAR'BAD'' and STARFLAGS = ''PERSIST-HIGH''  as well as telluric standard stars leaving us with a  total number  of 270 stars.  Those 270 include  5 known AGB stars and one supergiant star (VR 5-7)  in our sample. Table \ref{AGB} shows the stellar parameters for those stars.

From the 270 stars, AGB/supergiant candidates have been chosen as those brighter than $\rm H < 12.2$,  $\rm 1.3 < H-K < 3.7$, which excludes AGB/supergiant stars with strong IR excess,  and $\rm K_{0}= K - (1.82 * ( (H-K) - 0.2 )) < 4.7$, which is the dereddened magnitude  to put the star above the AGB tip. Note that the more massive ($>$3-4 solar masses) and O-rich AGB stars can be brighter (because of flux contribution by hot bottom burning) than the AGB theoretical luminosity limit (see e.g. \citealt{anibal2009}) and that to unambiguously distinguish between AGB or supergiant status one would need to have photometric light curves and/or some key chemical information from optical spectra (e.g. Lithium and s-process elements; \citealt{anibal2006,anibal2007}  (see also   \citealt{anibal2017} for a review). The factor 1.82 corresponds to the $\rm A_{K}/E(H-K)$ ratio from \citet{indebetouw2005}. 
The 15 AGB/supergiant candidate stars  in our final sample together with their stellar parameters are given in Tab.~\ref{super}.

\begin{figure}[!htbp]
  \centering
	\includegraphics[width=0.49\textwidth,angle=0]{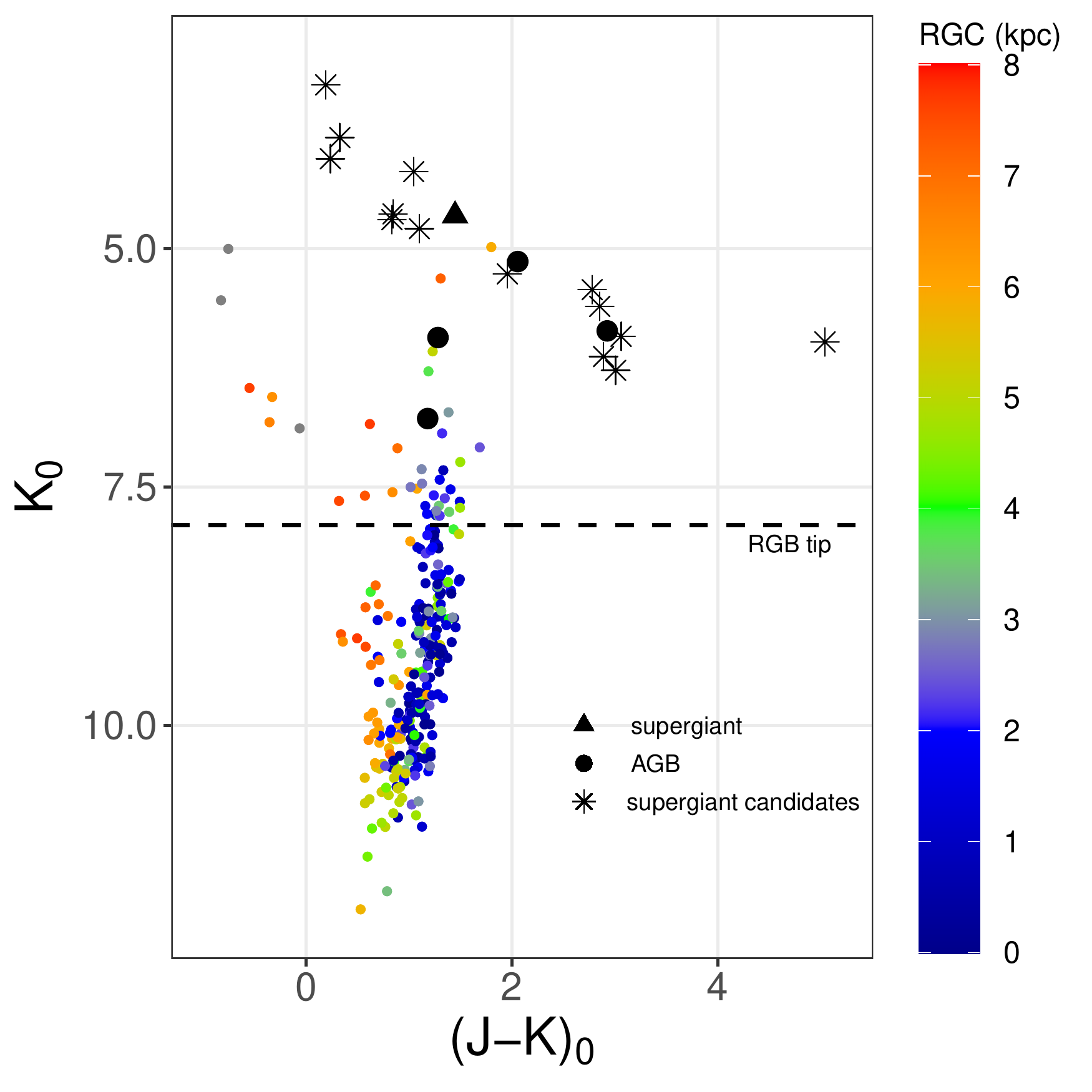}
	\caption{Dereddened color-magnitude diagram of the GALCEN sample as a function of the galactocentric distance. Superimposed are known AGB stars (filled circles), known supergiants (filled triangles)  and AGB/supergiant candidates (asterisks) based on photometric selection criteria (see text). } 
	\label{CMD}
\end{figure}

Figure~\ref{CMD} shows the dereddened color-magnitude diagram of the full GALCEN sample, together with the known AGB/supergiant stars and
candidate AGB/supergiant stars, colored by the galactocentric radius. As AGB stars and supergiants have circumstellar material due to their mass-loss, the dereddening of those stars were obtained by obtaining the median E(J-K) color excess from the  VVV extinction map (\citealt{gonzalez2012}) instead of the APOGEE estimate.
 Clearly visible is the foreground contribution with $\rm (J-K)_{0} < 0.8$ which can be easily removed by a simple distance cut.  The AGBs and supergiants (both known and candidates) are the most luminous stars and they lie well  above the tip of the RGB with $\rm K_{0}=7.9$  (\citealt{habing2004}), indicated by the dashed line. By imposing a cutoff in $\rm R_{GC} < 3.5\,kpc$  a total of 157 stars are left over for our analysis. We will refer to this and later on  as the {\it{GALCEN}} sample.

\begin{table}
\caption{Known AGB and supergiants with stellar parameters from APOGEE-DR16. Last column gives the references: (1) Schultheis et al. (2003), (2) Matsunaga et al. (2009), (3) Davies et al. (2009), (4) Cunha et al. (2007) }
\begin{tabular}{llllll}
2MASS ID&$\rm T_{eff}$&log\,g&[M/H]&Ref\\
  \hline
2M17451937-2914052&3730&0.34&-0.28&A58$^1$\\
2M17445261-2914110&3244&0.12&0.13&A20$^1$\\
2M17452187-2913443&3290&0.38&0.21&A62$^1$\\
2M17460808-2848491&3555&0.43&-0.46&V5009Sgr$^2$\\
2M17461658-2849498&3242&0.12&-0.15&VR 5-7$^{3,4}$\\

\hline
\end{tabular}
\label{AGB}
\end{table}

\begin{table}
\caption{Supergiant candidates}
\begin{tabular}{llll}
2MASS ID&$\rm T_{eff}$&log\,g&[M/H]\\
\hline
2M17460746-2846416&3584&0.95&0.29\\
2M17461382-2825206&3741&0.86&-0.24\\
2M17461772-2841159&3221&-0.02&-0.12\\
2M17462661-2819422&3218&-0.38&-0.26\\
2M17463072-2850325&3327&0.31&0.13\\
2M17463266-2837184&3692&0.55&-0.11\\
2M17463693-2820212&3346&0.90&0.13\\
2M17463769-2841257&3129&-0.15&-0.08\\
2M17464409-2817487&3128&-0.14&-0.05\\
2M17464864-2818274&3181&0.35&0.40\\
2M17471240-2838377&3536&0.31&-0.59\\
2M17472459-2822320&3337&0.03&-0.09\\
2M17472709-2840356&3360&0.23&-0.05\\
2M17480017-2821058&3292&0.23&0.13\\
\hline
\end{tabular}
\label{super}
\end{table}

\begin{figure}[!htbp]
  \centering
	\includegraphics[width=0.49\textwidth,angle=0]{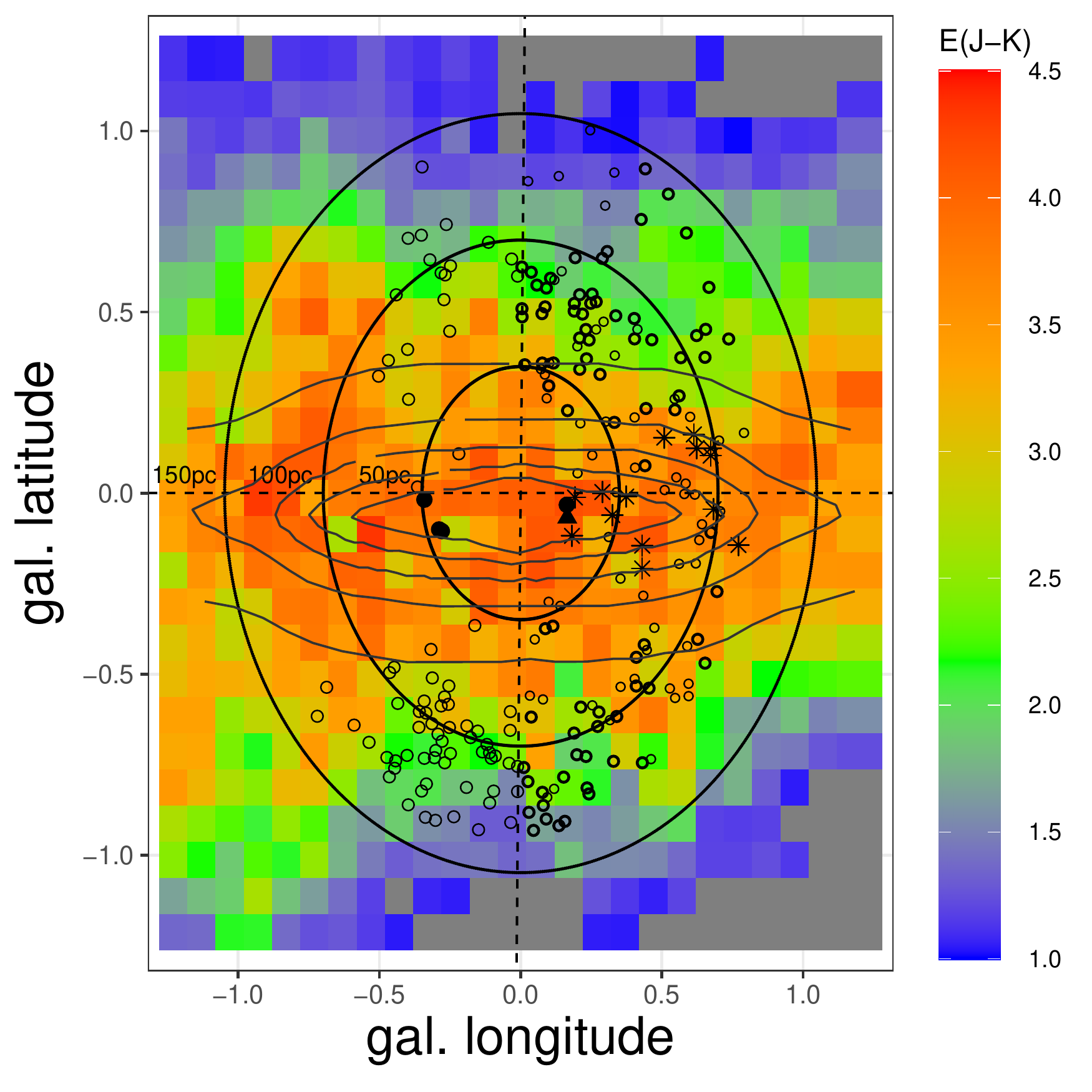}
	\caption{Galactic longitude vs. Galactic latitude distribution of the  GALCEN sample superimposed on the interstellar extinction map of \protect\citet{gonzalez2012}. The circles denote  radii of 50\,pc, 100\,pc and 150\,pc assuming a distance to the GC of 8.2\,kpc. Open circles denote M giant stars, the filled triangle is the supergiant VR 5-7, filled circles are AGB stars (see Tab. 1) and asterisks are supergiant candidates (see Tab. 2). The grey contours show the surface brightness map of the best-fit model of the ``nuclear bulge'' component by Launhardt et al. (2002).} 
	\label{coord}
\end{figure}

\begin{figure}
  \centering
        \includegraphics[width=0.49\textwidth]{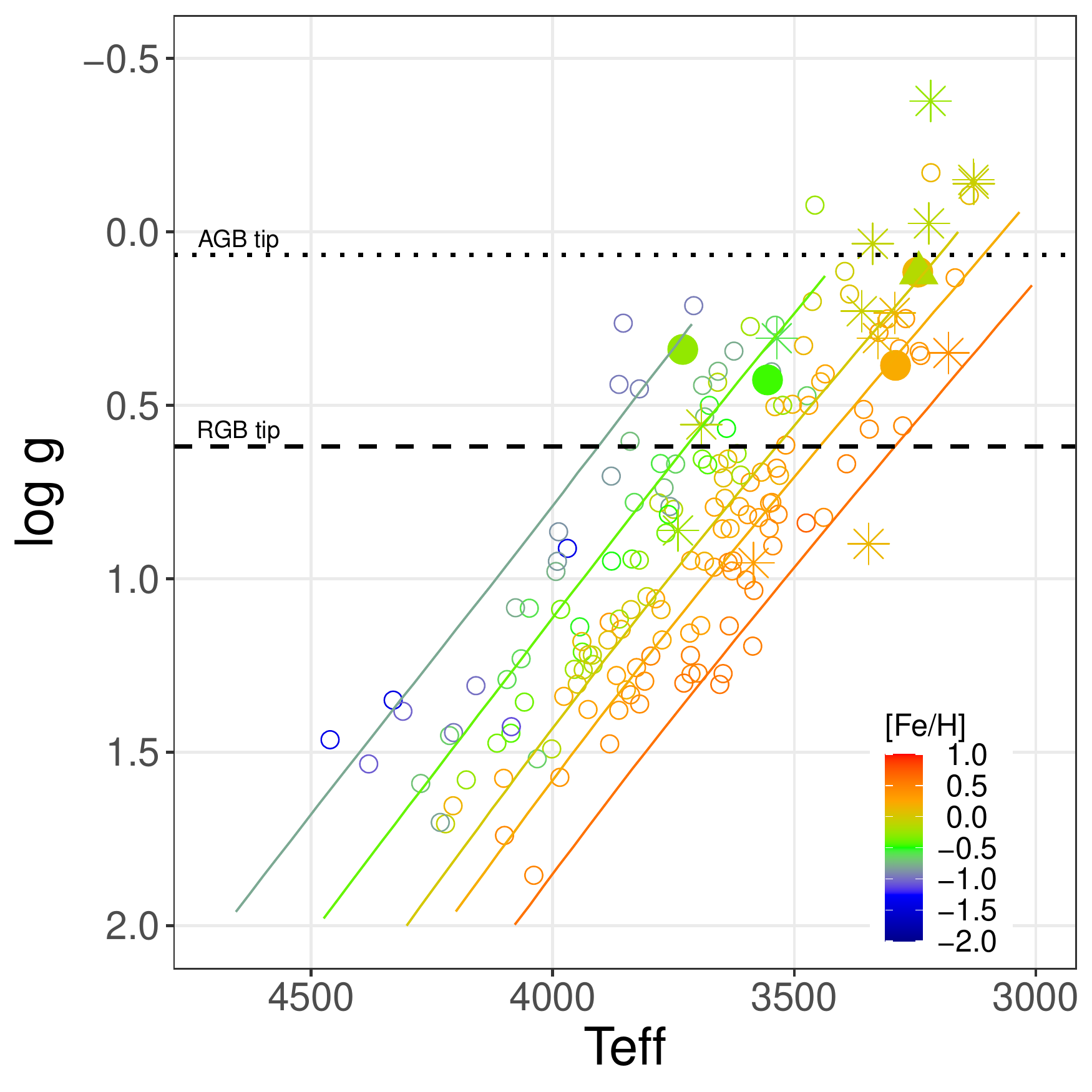}
	\caption{Teff vs. logg as function of   metallicity together with PARSEC isochrones  with an age of 8\,Gyr. Indicated are AGB stars as filled circles,
supergiants as filled triangles and supergiant candidates as asterisks. The dotted lines shows the approximate location of the tip of the AGB, while the dashed
line the tip of the RGB (see text).} 
	\label{stelparams}

\end{figure}

Figure~\ref{coord} shows the spatial distribution of the GALCEN sample in Galactic coordinates superimposed on the interstellar extinction map
of \citet{gonzalez2012}. In the highest extincted regions, mainly AGB stars (filled circles), supergiants (filled triangle) and supergiant candidates (asterisks) are located with projected distances closer than 50\,pc to  SgrA*. The majority of the M giants in our sample are located  in relatively low extinction windows  with $\rm E(J-K) < 3.0$ with projected distances of  50--150\,pc away from the GC. We also superimpose the stellar mass distribution of the nuclear disk from \citet{Launhardt2002}.

Stellar parameters and chemical abundances of up to 24 elements
are determined by the ASPCAP pipeline (see \citealt{garcia-perez2016}).
These values are based on a $\chi^{2}$ minimization between observed and synthetic model spectra (see \citealt{zamora2015}; \citealt{holtzman2015} for more details on the DR16 spectral libraries)
performed with the {\sc FERRE} code \citep[][and subsequent updates]{allende2006}.



Figure \ref{stelparams} shows the HR diagram of the  GALCEN sample colored by the metallicity. Superimposed are  PARSEC isochrones (\citealt{PARSEC}) assuming  an age of 8\,Gyr. Indicated  are the approximate limits of the RGB with $\rm M_{bol} < -3.5$ (\citealt{habing2004}) and the tip of the AGB  with $\rm M_{bol}=-7.0$ (\citealt{schultheis2003}). However, as pointed out by \citet{McQuinn2019}, the tip of the RGB becomes brighter for metal-rich stars with up to 0.3\,mag difference in the K-band (see their Fig. 6).
The spread in the effective temperature at a given surface gravity is mainly due to the metallicity, nicely also indicated by the isochrones. The known AGB stars are situated above  the tip of the RGB and show similar metallicities as expected from their location in the HR diagram following the stellar isochrones. We notice, however, two stars, 2M17462584-2850001 and 2M17470135-2831410, showing very low metallicities ($\rm [Fe/H] < -2.0$) where their projected metallicities on the HR diagram do not match the isochrone metallicities. Their ASPCAP fits are particularly bad (i.e. relatively high $\rm \chi^2$ values). One of these stars (2M17462584-2850001) is a known AGB star with a Mira-like long variability period of 519 days, while the other star (2M17470135-2831410) is an AGB/supergiant candidate with no period information available but with a near-IR variability more typical of an extreme AGB star. The automatic pipeline ASPCAP does not work well for extreme AGB/supergiant stars (e.g. the more massive and/or evolved dusty AGBs or extreme and dusty red supergiants; APOGEE/ASPCAP team, priv. comm.). For the very few known extreme Galactic disk (solar metallicity) O-rich AGB stars observed by APOGEE, ASPCAP systematically gives much lower metallicities (even by ~2 dex) for the longer period stars, which are expected to be the more massive and/or evolved dusty AGBs. There are several reasons for this like veiling by hot dust emission, stronger degeneracies due to their much more complex spectra and/or the specific pulsational phase during the observations, with the first one being very likely the dominant factor. This is because the strong hot dust emission in these stars veils the spectrum and the molecular bands look much weaker than reality; ASPCAP could thus compensate this with a much lower global metallicity and/or a higher Teff. Thus, we are not confident about the ASPCAP results for these two particular and extreme stars.


\section{AGB stars and supergiants}
\subsection{Previously known AGB  stars and supergiants}

\citet{schultheis2003}[hereafter referred to as S03]  performed low-resolution spectroscopic follow-up observations of sources with bright excess at 7 and 15$\rm \mu m$   from the ISOGAL survey (\citealt{omont2003}), which probed the stellar populations of the inner region of the Milky Way. The majority of these sources are long-period variables on the AGB with strong mass-loss, well traced by their 15\,$\rm \mu$m excess (\citealt{glass1999}). S03  have shown that
the molecular bands of $\rm ^{12}CO$ and $\rm H_{2}O$, together with the bolometric magnitudes, are an excellent indicator  of  stellar populations such as AGB stars, supergiants, red giants and young stellar objects.

\noindent
\begin{itemize}
\item  2M17451937-2914052  is a high luminosity  OH/IR star from the OH/IR star sample of \citet{ortiz2002}. These are the most extreme AGB stars, with large pulsational periods (several hundreds of days) and mass-loss rates up to a few  times $\rm 10^{-5} M_{\odot}/yr$,  displaying the highest bolometric luminosities ($\rm M_{bol} < -4$). S03 estimated the $\rm M_{bol} \sim -5.13$ and the mass-loss rate of  $\rm 8\times 10^{-6} M_{\odot}/yr$. The Near-IR low-resolution spectrum of S03 shows extremely strong water absorption at about 1.7\,$\mu$m, typical for large amplitude pulsation (see e.g. \citealt{lancon2000}).

\item  2M17462584-2850001 is also a known OH/IR star which was monitored by \citet{wood1998}  and
appears also in the catalog of large amplitude variables of \citet{glass2001} and \citet{matsunaga2009}. It has a period of 505 days (while 519 days in \citealt{matsunaga2009}) and an amplitude in K of 1.60\,mag. It has a very low ASPCAP metallicity of $\rm [Fe/H] \sim -2.17$  compared to the others. The ASPCAP $\chi^{2}$ value is higher than 100,  compared to a typical value of about 20--50  for the other  stars,  making the stellar parameters, including the global metallicity, unreliable. We omit this star for our study.

\item 2M17445261-2914110 and 2M17452187-2913443 are classified by S03 as AGB star candidates with  $\rm M_{bol} = -4$ and -4.77, respectively. They have moderate mass-loss rates of $ \sim 7.4 \times 10^{-7}$ and $\rm 2 \times 10^{-7}   M_{\odot}/yr$, respectively. They both also show some moderate water absorption.

  \item 2M17460808-2848491 is  a large-amplitude Mira Variable, discovered first by \citet{glass2001} and reconfirmed by \citet{matsunaga2009}. It has a period of 297 days with
 an K amplitude of  0.55\,mag.

\item 2M17461658-2849498  (VR 5-7) is a member of the Quintuplet Cluster (\citealt{Moneti1994}) and has been classified as a supergiant with a spectral type of M6I (\citealt{liermann2009}).
\citet{cunha2007} obtained a photometric temperature (based on the measurements of the CO molecular band)   of 3600\,K,   a photometric log\,g of -0.15\,dex, and an iron abundance of +0.14\,dex.
 \citet{Davies2009}  observed VR 5-7 in H-band
at Keck with NIRSPEC and a spectral resolution of 17,000. They obtained a $\rm T_{eff}$ of $\rm 3400 \pm 200\,K$, a log\,g of $\rm 0.0 \pm 0.3\,dex$ and  a $\rm [Fe/H]=0.10 \pm 0.11\,dex$. While the temperature and 
surface gravity agrees within the errors, the  metallicity of APOGEE is about  0.2\,dex  lower. 

\end{itemize}

\begin{figure*}
  \centering
        \includegraphics[width=0.99\textwidth]{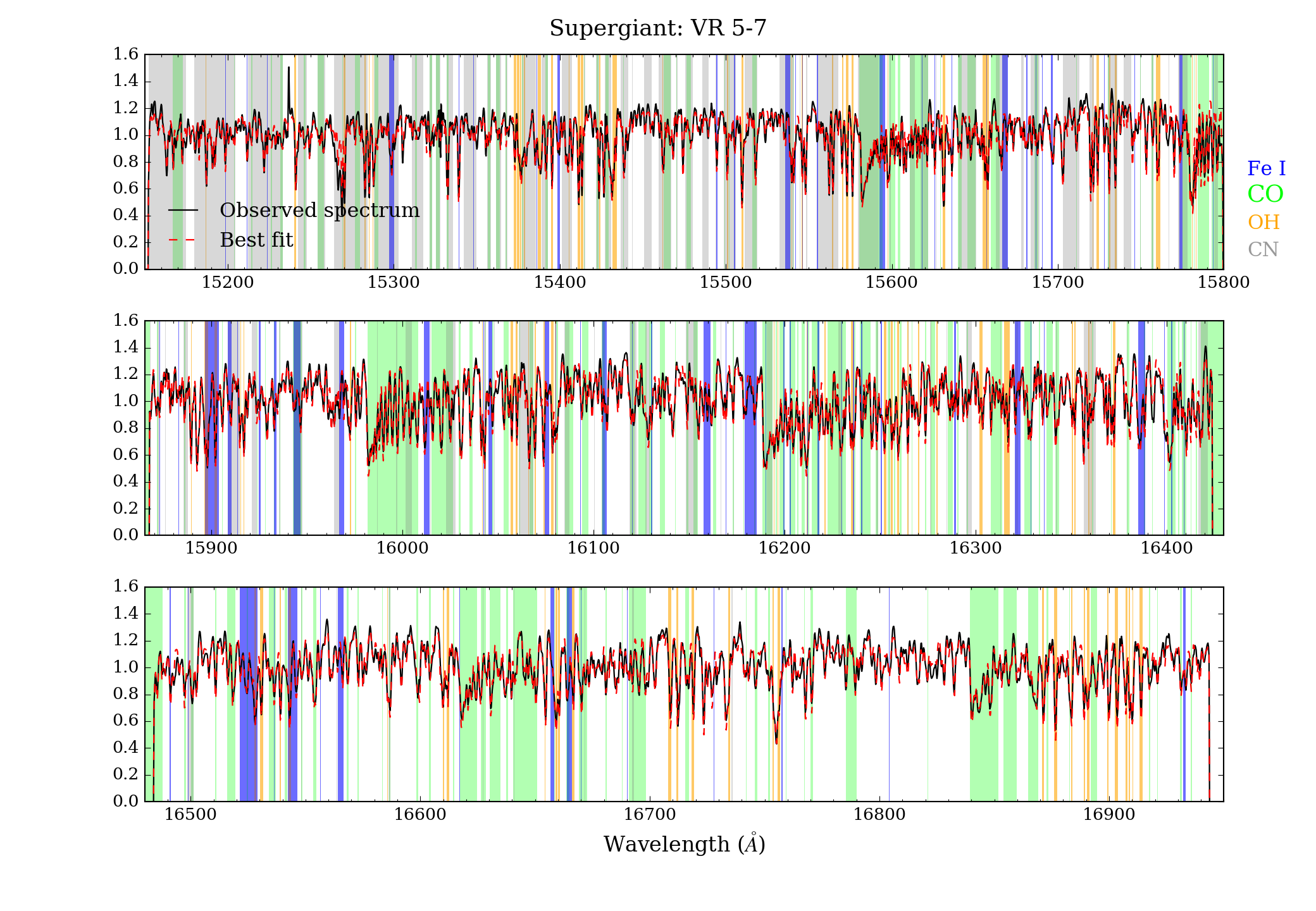}
	\caption{APOGEE spectrum of the supergiant star VR5-7. Black shows the observed spectrum and red the best fit spectrum obtained by the automated APOGEE pipeline ASPCAP. The most prominent molecular bands of CO, CN and OH as well as Fe I lines are also indicated.} 
	\label{spec}

\end{figure*}

\begin{figure}[!htbp]
  \centering
 \includegraphics[width=0.49\textwidth]{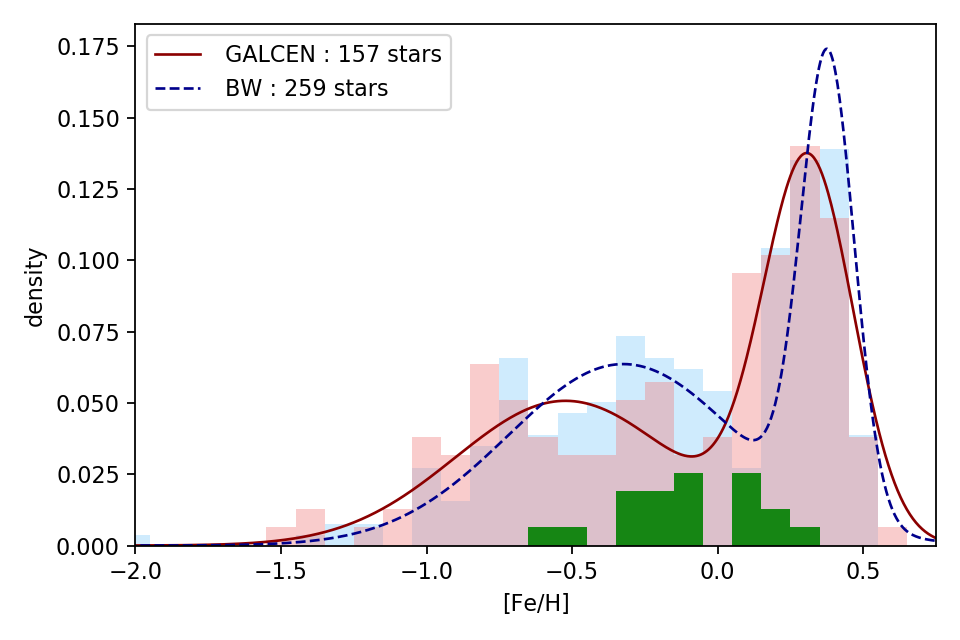}
     \caption{GMM decomposition of the MDF for the GALCEN sample (red)  and Baade's window (blue). In green color   the known and candidate AGB/supergiant stars (Tables 1 and 2) are indicated. The bin size is 0.1\,dex.}
       \label{histFeHGALCEN}
        \end{figure}

Figure~\ref{spec} shows the APOGEE spectrum of VR 5-7 together with the best model fit of ASPCAP (in red). We clearly see the strong impact of the molecular lines such as CN, OH and CO (indicated as gray, orange and lime, respectively)
\subsection{AGB/supergiant candidates}
As mentioned in Sect 2.3, AGB/supergiant candidates have been chosen using a  photometric  color-cut in (H--K) as well as the dereddened magnitude
cut of $\rm K_{0}$ to put the star above the tip of the AGB.   In total 15  AGB/supergiant candidates are in our sample with the stellar parameters given in Tab.~\ref{super}.
Twelve stars out of fifteen lie indeed above the tip of the RGB in Fig. ~\ref{stelparams} and five of them even above the tip of the AGB. One of them,
2M17463266-2837184,  shows an extreme $\rm (J-K)_{0}$ color of about 5  and could be an AGB star with strong mass-loss. The three remaining stars are only about 0.2--0.3\,dex below the tip of the RGB   which is  approximately the  uncertainties of the derived surface gravities  of APOGEE for those kind of objects.
This shows that a photometric H--K cut together
 with the dereddened K magnitude is a priori a good first indication of detecting M AGB/supergiants. 
 Three additional stars lie above the tip of the AGB which show also very high luminosities in the  CMD (Fig.~\ref{CMD}) making them excellent candidates of being supergiant stars. 
The relatively high fraction of supergiants in GALCEN is compatible with a recent increased star formation rate during the last 200-300 Myr (\citealt{Pfuhl2011}).

\section{Metallicity distribution function}

\citet{Schultheis2015}  already had  some indications for the existence of a metal-poor population in GC region based on their metallicity distribution function (MDF) as shown in  their Fig. 4.
With our larger sample here, we performed a Gaussian mixture modeling (GMM) decomposition, which is a parametric probability density function given by the weighted sum of a number of Gaussian components. In order to constrain the number of Gaussians, we adopted the Akaike information criterion (AIC), which gives preference of a two-component solution. For comparison we also performed a GMM for Baade's window (BW) with the same sample criteria,  defined as in \citet{schultheis2017} but applied to the DR16 data release.

%

We find in both samples, GALCEN and BW, a narrow metal-rich component and a broader metal-poor component. The metal-rich component
of both fields are centered at around $\rm [Fe/H]= +0.3\,dex$, with a slightly  larger dispersion for the GALCEN field. Very striking and visible is
the metal-poor peak in the GALCEN field centered at $\rm [Fe/H]=-0.53\,dex$. This is about 0.2\,dex more metal-poor than the equivalent peak in BW while the dispersions are  very similar ($\rm \sim 0.4\,dex$). 63\%   of the population in  BW is  in the metal-poor regime while for the GALCEN field it is about 47\%
(73 stars out of 157).

In order to test the  statistical significance we performed a  Bootstrapping analysis with 1000 resamplings. We find for BW the metal-rich peak at $\rm [Fe/H]= -0.33 \pm 0.03$ and for  the metal-poor component  a $\rm [Fe/H]= 0.38 \pm 0.01$ while for GALCEN the two components are situated at $\rm [Fe/H]= 0.30 \pm 0.03$ and $\rm [Fe/H]= -0.54 \pm 0.08$, respectively.
    The mean weight of the metal-poor component of  BW is $\rm 62.8 \pm 3.6\%$ while for GALCEN it is $\rm 46.6 \pm 6.3\%$ leading to a statistically significant  difference of 2.23 sigma.

 Indicated in green are also the previously known 4 AGB stars and the supergiant star VR 5-7, as well as the AGB/supergiant candidates.  A visual inspection of the overall shape of the MDF of these cool stars shows that  they follow the general trend of the normal red giants. This makes us confident that indeed the metallicities of those cool stars are reliable. Unfortunately, the sample size is too small to perform any statistical test. We also notice that the dispersion in the metallicity for these stars is much narrower ($\rm \sim 0.25\,dex$) compared to the  M giants, however clearly a larger sample size is needed in order to attribute differences in the MDF.

\begin{figure*}
\centering
\includegraphics[width=0.99\textwidth,angle=0]{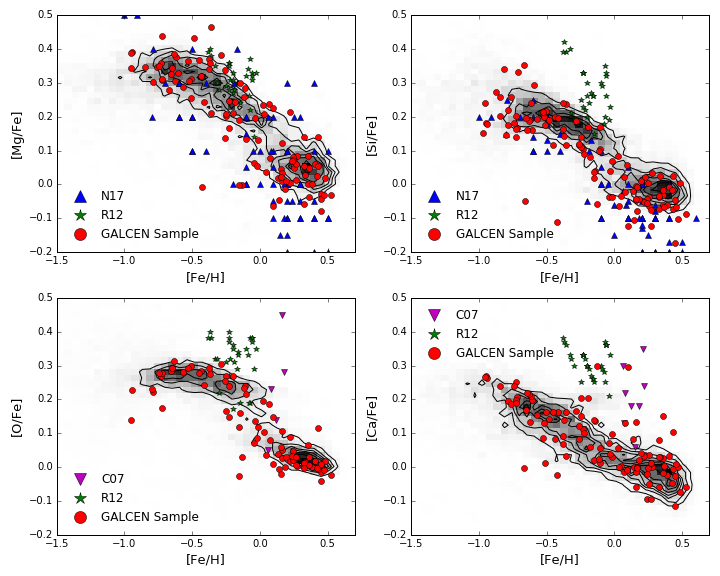}

\caption{[Fe/H] vs [X/Fe] of the alpha-elements, for the APOGEE GALCEN sample.  Also shown for comparison are the APOGEE bulge sample of Queiroz et al. (2020, in prep) as contours, the APOGEE bulge sample of Zasowski et al. (2019, updated for DR16) as shaded density, and literature values for Cunha et al. (2007, C07), Rich et al. (2012, R12), and Nandakumar et al. (2017, N17).}
\label{abundancesalpha}
\end{figure*}

\citet{zoccali2017} used  simple limits of $\rm [Fe/H]>+0.1$ to define metal-rich stars, and $\rm [Fe/H]<-0.1$ for the metal-poor ones, to argue that the metal-rich population of
their GIBS data  is flattened and concentrated towards the Galactic plane, while the metal-poor component is spheroidal. \citet{schultheis2019} used these same criteria  to detect
a very prominent metallicity spike (see their Fig.~4) in the Galactic Center (including the Nuclear Star Cluster), where  about 80\% of their sample is metal-rich. If we apply the same criterion to the GALCEN field, we find about 50\% of the stars at $\rm |z|=0.80\,kpc$ to be  metal-rich, in perfect agreement within the inner four degrees as shown in Fig.~4 of \citet{schultheis2019}. This strengthens the evidence for  a distinct stellar system in the GC region with respect to the Inner Galactic Bulge ($\rm |b| < 2^{o}$).

\section{Chemical Abundances}

\citet{Schultheis2015} could only study
the general $\alpha$-element abundances for the GALCEN field with APOGEE DR12,  but with DR16 we are able to study the detailed chemistry of the cooler end of the giant branch. We restrict here our analysis to regular M giant stars and  do not include the known and candidate  AGB/supergiants stars. As a comparison sample, we use the Galactic Bulge sample by Queiroz et al. (2020, in prep.) based on SH distances and the \citet{zasowski19} Bulge sample, updated for DR16. The sample was defined  as stars falling inside the Galactocentric coordinates $\rm |X| < 5\,kpc$, $\rm |Y| < 3.5\,kpc$, $\rm |Z| < 1.0\,kpc$ with an additional cut in the
 reduced proper motion diagram. We refer here to Queiroz et al. (\citealt{queiroz2020}) for a detailed description on their Bulge sample.
 In addition, we restricted our sample to $R_{\rm GC} \le 3.5$~kpc., leaving about 5800 Bulge stars for our comparison which is shown in Fig.~\ref{abundancesalpha}. 
 The detailed properties of the comparison Bulge sample will be discussed in an accompanying paper. In addition, we show as a comparison the Bulge sample of \citet{zasowski19}, updated for DR16.


We use calibrated chemical abundances and we omit stars where the individual error on each studied chemical element is larger than 0.02\,dex. The overall behaviors for the alpha elements Mg, Si, O and Ca are shown as a function of [Fe/H] in Fig.~\ref{abundancesalpha}.
Due to the large  number of bulge comparison stars, we use densities for the \citet{zasowski19} sample and  contours for the Queiroz et al. sample
(see  Fig.~\ref{abundancesalpha}). 
In addition, we use samples in the literature of \citet{rich2012}, \citet{Nandakumar2018} in the inner Bulge ($\rm |b| < 1^{o}$)
and \citet{cunha2007} for stars in the nuclear star cluster.

The GALCEN stars (black dots) overall follow the same behavior as the Galactic Bulge stars of Queiroz et al. (2020, in prep.).
Concerning the behavior of APOGEE results for [$\alpha/Fe$]--[Fe/H] it is clear that Ca shows, compared to the other $\alpha$-elements, a different behavior where stars with subsolar metallicities are less enhanced compared to Mg, O and Si; the Ca abundances
also decrease for higher metallicities while the other alpha-elements tend to show a flattening. In the DR16 results, there is a noticeable feature (reminiscent of a ``finger", J\"onsson et al. 2020) at $\rm [Ca/Fe] \sim +0.25$ and  $\rm [O/Fe]  \sim +0.25$ and  $\rm [Fe/H] > 0$; such horizontal sequence of stars
had already been noticed for the DR14 results in the inner Bulge sample (\citealt{zasowski19}).  J\"onsson et al. (\citealt{Joensson2020}) concluded that the tightness of this feature, which contains about 4\% of the main stellar sample of giants, makes it likely that this feature is due to some possible systematics in the abundance determinations, which the APOGEE/ASPCAP team has not yet been able to identify.
However, most likely due to our small sample size, we do not find this feature in the GALCEN sample.

As in DR14, we see that the oxygen abundances show the tightest sequence compared to Mg, Si and Ca, in particular for low metallicities. Worthwhile mentioning is the extremely flat behavior of oxygen for high metallicities ($\rm [Fe/H] > 0.1$), as already noted by \citet{zasowski19}; the DR16 data release using the MARCS models confirm this trend. However, \citet{Johnson14} found in their analysis of Bulge stars in the optical and by comparing them
with values in the literature that the O-abundances decrease for higher metallicities with no sign of a flattening (see their Fig. 13). So this flattening at
higher metallicities found in the APOGEE data has to be taken with caution as it could be an artifact; however, deriving oxygen abundances for high-metallicity stars in the optical can also be challenging. For example, very recently \citet{pablo2020}  has shown based on high-resolution optical spectra, that this flattening for  the metal-rich stars can be explained by an incorrect continuum normalization. They demonstrate that with a  proper continuum normalization,  for example the alpha-element Mg decreases steadily with increasing metallicity with no signs of a flattening.
From the theoretical point of view, the flattening can be reproduced  by increasing the stellar yields for high metallicities by a factor of 3.5 (see \citealt{Matteucci2020}) or by radial migration (see e.g \citealt{Minchev2014}, \citealt{Anders2017}). 
Si and Mg show very similar trends with a less pronounced knee position than  oxygen. Striking  for both elements is the flattening at high metallicities while \citet{Nandakumar2018} show a decreasing trend of Si and Mg for the metal-rich stars. \citet{Johnson14} shows for their Bulge sample that Mg and Si decrease as well with increasing metallicity.
Interestingly, the sample  of Cunha et al. (2007),  which are stars in the nuclear star cluster, are also enhanced, but in a single metallicity.

\section{Is  the nuclear star disc and the  nuclear star cluster chemically distinct?}
\begin{figure}
\centering
\includegraphics[width=0.49\textwidth,angle=0]{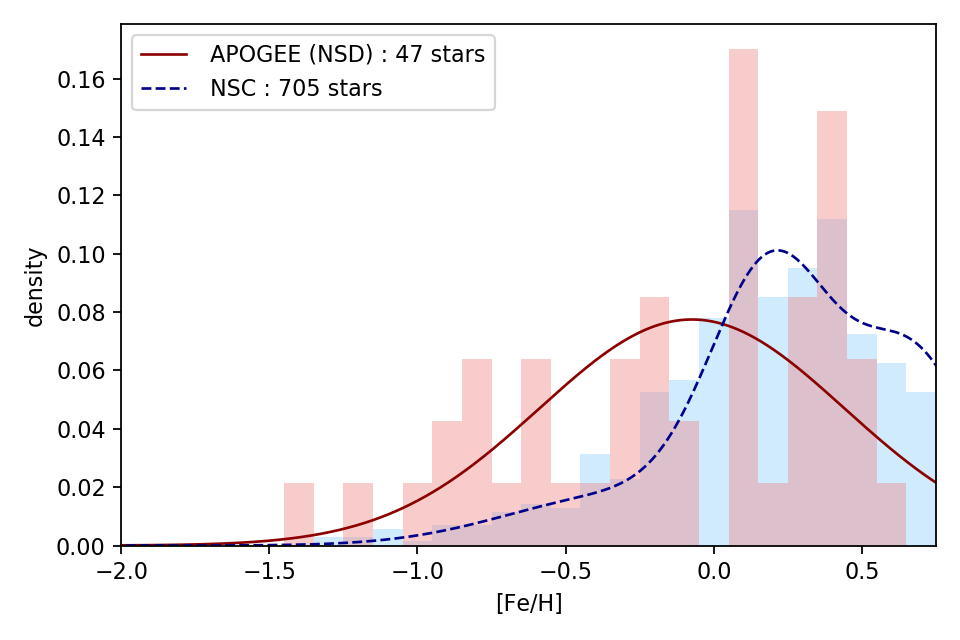}
\caption{MDF of the NSC sample of Feldmeier-Krause et al. (2020)  compared to the NSD GALCEN sample which is confined to  $\rm |b| < 0.4^{0}$}

\label{MDFNSD}
\end{figure}

\citet{cunha2007}  already  noticed for their  sample of M giants in the NSC  $\rm \alpha$-enhancement in $\rm [O/Fe]$ and $\rm [Ca/Fe]$ at super-solar metallicities.
Recently,  \citet{Thorsbro2020} reveals also alpha-enhanced, metal-rich stars in the NSC which could be sign of a starburst activity, initiated by a sudden accretion of gas, either primordial or slightly enriched.
Although their sample is small and clearly more stars are needed to confirm their results, they  show a possible sign of a  starburst,  making
a kind of ``loop'' structure in the $\rm [Si/Fe] vs. [Fe/H]$ diagram which can be explained as following: The sudden accretion of gas will first result in a dilution
of chemical abundances (e.g. Fe) before the $\rm [Si/Fe]$ vs.$\rm  [Fe/H]$ ratio increases due to the contribution of core-collapse SNe II, following by a decrease of
$\rm [\alpha/Fe]$ due to the Fe production from Type Ia SNe. In our Fig.~\ref{abundancesalpha}, which probes the nuclear stellar disc (NSD)  we do not see evidence of  alpha-enhanced metal-rich stars suggesting  a chemical distinction between the NSC and the NSD.  In addition, \citet{Feldmeier-Krause2020} traced the metallicity distribution function in the NSC based on 600 late type stars. We compare the MDF in Fig.~\ref{MDFNSD} between their sample and our GALCEN sample which we restrict to  $\rm |b| < 0.4^{0}$   in order to probe the NSD  (see Fig.~\ref{coord}).  Although only $\sim$ 50 stars are left over to be
in the NSD, we see that the two MDFs differ, in particular  the MDF of the NSD shows a more significant metal-poor tail ($\rm [Fe/H] < -0.5$) compared to the NSC. Performing a bootstrapping with 1000 resamplings we find for the NSD the metal-poor peak at $\rm -0.57 \pm 0.19\,dex$, while for  the NSC at $\rm -0.36 \pm 0.12\,dex$. However, the sample size of the NSD is clearly too small and more observations are needed.

In addition, we see also differences in their  star formation histories.  While the majority of stars in the NSC and NSD are old ($\rm 80\%$ of the stars are $\sim$ 10\,Gyr old), \citet{nogueras2020}  detected sign of a recent  star formation around 1 Gyr ago in the NSC which is not present for the NSC (\citealt{Pfuhl2011}, \citealt{Schoedel2020}). This strengthens the argument that indeed the NSC and NSD are chemically distinct and these two systems might have formed differently.

\section{Kinematics versus metallicity} \label{kinematics}

We transformed the heliocentric radial velocities to galactocentric velocities according to \citet{schoenrich2015}:

\begin{equation}
   \label{equation}
   v_{GC} = v_{rad} + 14\times cos(l)cos(b) + 250\times sin(l)cos(b) + 7\times sin(b).
\end{equation}

\begin{figure}[!ht]
\centering
\includegraphics[width=0.45\textwidth,angle=0]{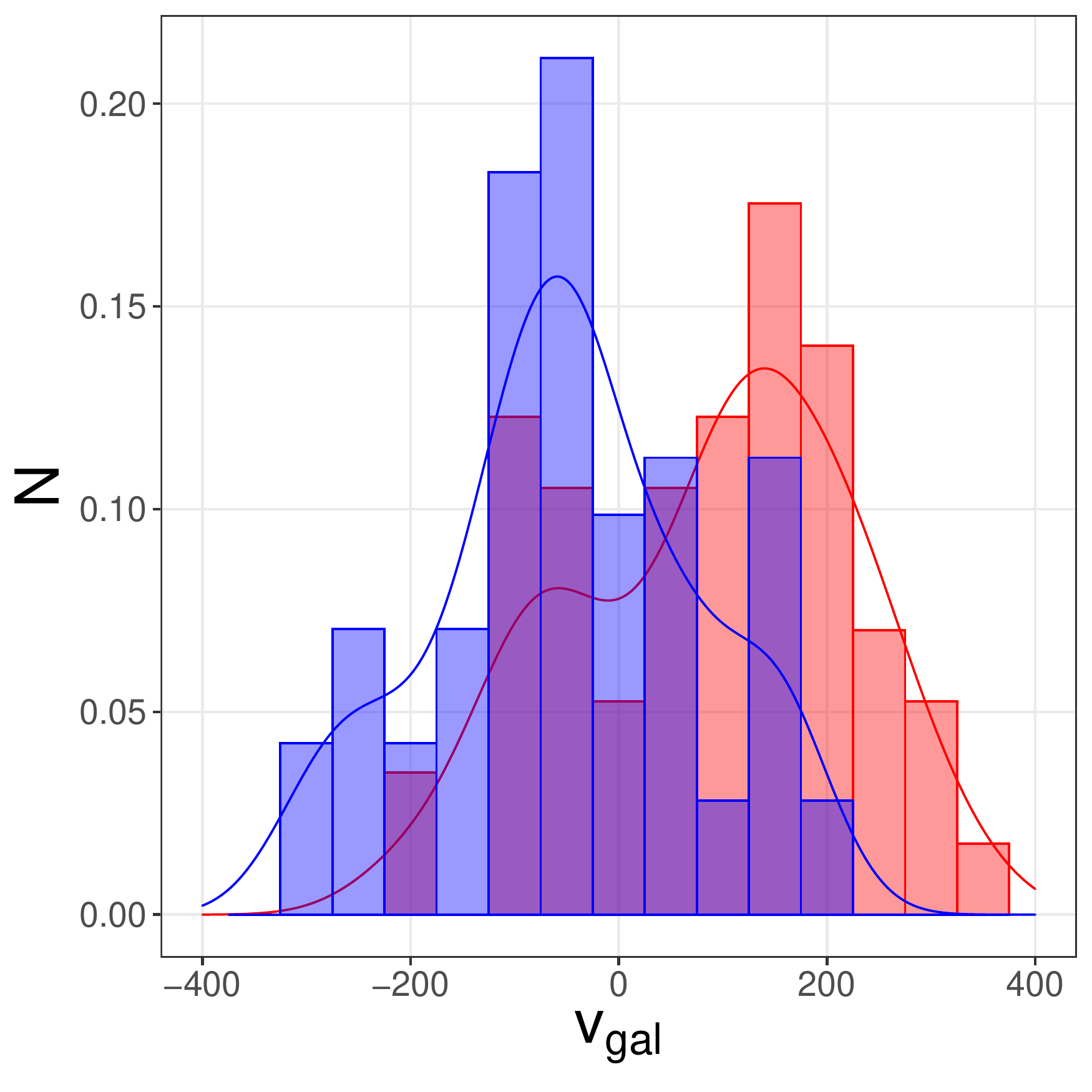}  

\caption{Normalized galactocentric velocity distributions for negatives longitudes and positive longitudes (in red and blue, respectively) as in \protect\citet{schoenrich2015}. We limit here our sample to $\rm |l| < 0.5^{o}$ in order to account for  the uneven sample distribution. }
\label{rotation}
\end{figure}

Figure~\ref{rotation} shows the normalized  galactocentric velocity distribution split in negative and positive longitudes where we confirm the clear signature of the
rotation of the nuclear stellar disk as pointed out by \citet{schoenrich2015}.  
However, due to the larger sample size now in GALCEN, we noticed a significant fraction of high velocity stars exceeding 300\,km/s.  These high velocity stars have $\rm [Fe/H] > -0.5$ ruling
out any possible connection to the Galactic Halo. Are these stars originating from the Galactic Center, or could they have been kicked up from e.g. binary supernova explosion, the interaction of a dwarf galaxy or a globular cluster with the disk, or interaction between multiple stars, as proposed by \citet{Du2018}? Proper motion measurements would be necessary  to derive the detailed orbits of those stars in order to understand their origin.

\begin{figure}[!ht]
\centering
\includegraphics[width=0.45\textwidth,angle=0]{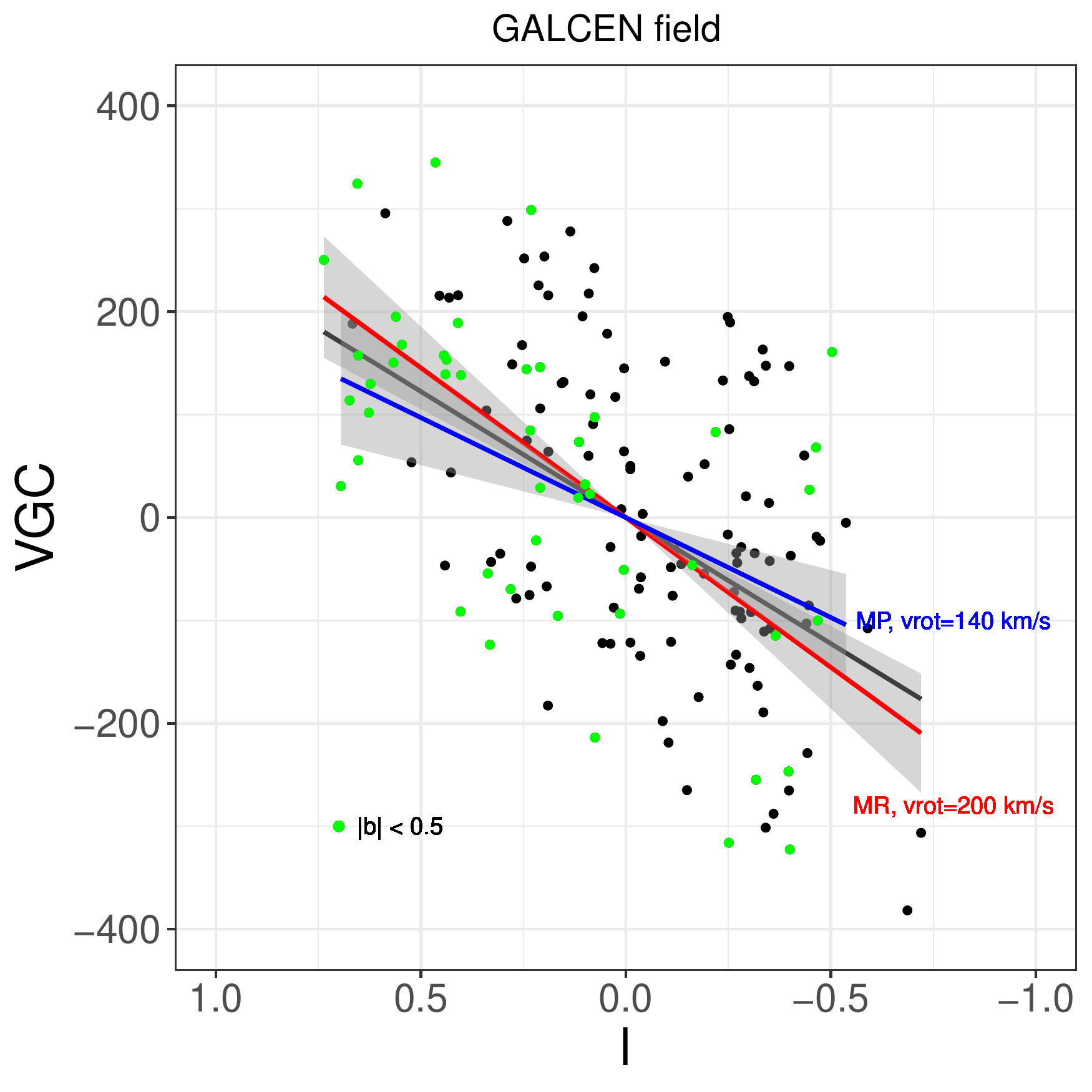}  

\caption{Galactocentric radial velocity vs. galactic longitude.  The black line shows the linear fit of the full GALCEN sample, while the red line the fit for metal-rich stars $\rm [Fe/H] > 0$ and the blue line for MP stars. The shaded area shows the 95\% level of confidence interval.  The green points are stars  inside 0.4 degrees gal. latitude. See text for details. }
\label{rotationMH}
\end{figure}

  The velocity dispersion of the GALCEN sample is 159\,km/s, with a slightly larger dispersion for the metal-rich stars (163\,km/s) with respect to the metal-poor stars (156\,km/s).  We define here metal-rich  (MR) and metal-poor (MP)   as in \citet{zoccali2017} with $\rm [Fe/H] > 0.1$ for MR and $\rm [Fe/H] < -0.1$ for MP. 
  \citet{zoccali2017} showed based on GIBS data that while for higher latitudes the MP population shows a larger velocity dispersion, this trend is inverted by going closer to the plane from $\rm |b|=1^{o}$ inwards. \citet{Clarkson2018} found based on proper motion measurements   that the metal-rich population shows a steeper rotation curve.  Our analysis of GALCEN confirms their results with an increasing velocity dispersion with higher metallicity. \citet{Kunder2016} showed that metal-poor stars such as RR Lyrae rotate much slower than RGB or RC stars in the Bulge and suggested that they  are belonging to the classical bulge.  Other studies also reveal the presence of an old population in the Galactic Center region (see e.g. \citealt{Contreras-Ramos2018}, \citealt{Dong2017}, \citealt{Minniti2016}).
  \citet{zoccali2017} showed some indications based on the  GIBS RC stars that MP stars show a marginally slower rotation compared to the MR stars (see their Fig. ~13). We investigated this more in detail by tracing the slope of the MR and MP stars in the
  $\rm l$ vs. $\rm v_{GC}$ plane. We use a similar way in the fitting procedure as
  \citet{schoenrich2015}. Figure~\ref{rotationMH} shows a linear fit of the entire GALCEN sample (black line) as well as the fit for the MR stars (red solid line) and MP stars (blue solid  line). The rotational speed can be calculated by assuming a radius for the NSD as shown by
  equation (\ref{equation2}) assuming a typical radius R  of the NSD of 120 pc.


\begin{equation}
   \label{equation2}
   v_{rad} = R_{0}\times sin(l) \times (V/R_{NSD} - V_{0}/R_{0})
\end{equation}


where $\rm v_{rad}$ is the heliocentric radial velocity, $\rm R_{0} =8.2\,kpc$ (distance to the GC), $\rm V_{0} =220\,km/s$ (sun velocity),  l the galactic longitude and $\rm R_{NSD}$ the radius of the disc (120\,pc).

We obtain  for the best fit (indicated by the blue and red line for both the MP and MR population, respectively)  a rotation velocity  of $\rm 140 \pm 30\, km/s$  for the MP stars and of $\rm 200 \pm 30\,km/s$ for the MR stars, indicating a clearly  faster rotation for the MR stars. If we restrict our sample to $|b| \le 0.4^\circ$, which is the approximate vertical extent of the nuclear stellar disc (see the Fig.~2 ), we obtain very similar fitting results.
The reason of this difference in the rotation velocity between MR and MP stars could be due to a different origin, e.g. MP stars  could be formed from disrupted stellar clusters. \cite{Tsatsi2017} showed that using N-body simulation of inspiralling clusters to the center of the Milky Way can reproduce both the morphological and kinematical properties of the Nuclear Star Cluster.
On the other hand, \citet{alessandra2019} has shown that, in presence of an aspherical, time-variable potential (e.g. due to the presence of a discrete stellar black hole cusp), counter rotating discs can perturb the dynamics of older stellar populations leading to accelerated relaxation while the younger populations show a larger amount of rotation.
(Younger populations always show larger rotation compared to older ones, but the difference is even larger for counter rotating discs).
Age information and chemical footprints  such as dispersion in light elements (\citealt{Schiavon2017}) would be essential  to understand the different possible formation scenarios.



\section{Conclusions}

Using the latest DR16 APOGEE data release, we study cool M giants in the GALCEN sample going down to temperatures of 3200\,K.  Among those, stellar parameters of a bunch of known  and candidate AGB/supergiant stars are now available thanks to the cool grid of MARCS model atmospheres and significant improvements in the analysis of cool stars by the APOGEE/ASPCAP team. The known (6) and candidate (15) AGB/supergiant stars are situated well above the tip of the RGB, confirming their status as cool luminous stars.  We confirm that a photometric H--K color cut combined with a  dereddened magnitude cut of $\rm K_{0}$ is a very powerful criterion to select AGB/supergiant stars. We identify clearly in the MDF of GALCEN a metal-poor peak at $\rm [Fe/H]=-0.53\,dex$, which is about 0.2\,dex  lower than the metal-poor peak in BW. About 50\% of the stars belong to each of the MR  and MP  populations. The group of AGB/supergiant stars follow the same trend in the MDF as the normal M giants.
The general $\alpha$-abundances of the GALCEN stars show a very bulge like behavior. As already pointed out by \citet{schoenrich2015}, we identify the rotation of the nuclear disc seen in the galactic longitude vs. $\rm v_{GC}$ diagram. Separating our sample in an MR and MP population, we detect a much higher rotation velocity of the MR stars with $\rm 200\,km/s$ with respect to the MP stars ($\rm 140\,km/s$), indicating a different origin of these two populations. We see evidence of clear differences in the chemical footprints of the nuclear disc and the nuclear star clusters, both in chemical abundances and the MDF. Together with their different SFH,  it is likely that the NSD and the NSC have formed on different time scales.

\begin{acknowledgements}
  We want to thank the anonymous referee for her/his very fruitful and valuable comments.
  MS  acknowledges the Programme National de Cosmologie et Galaxies (PNCG) of CNRS/INSU, France, for financial support. DM is supported by the BASAL Center for Astrophysics and Associated Technologies (CATA) through grant AFB 170002, by the Programa Iniciativa Científica Milenio grant IC120009, awarded to the Millennium Institute of Astrophysics (MAS), and by Project FONDECYT No. 1170121. DAGH and OZ acknowledge support from the State Research Agency (AEI) of the Spanish Ministry of Science, Innovation and Universities (MCIU) and the European Regional Development Fund (FEDER) under grant AYA2017-88254-P.  J.G.F-T is supported by FONDECYT No. 3180210 and Becas Iberoam\'erica Investigador 2019, Banco Santander Chile. S.H. is supported by an NSF Astronomy and Astrophysics Postdoctoral Fellowship under award AST-1801940. R.R.M acknowledges partial support from project BASAL AFB-$170002$ as well as FONDECYT project N$^{\circ} 1170364$.

Funding for SDSS-III has been provided by the Alfred P. Sloan Foundation, the Participating Institutions, the National Science Foundation, and the U.S. Department of Energy Office of Science. The SDSS-III web site is http://www.sdss3.org/. SDSS-III is managed by the Astrophysical Research Consortium for the Participating Institutions of the SDSS-III Collaboration including the University of Arizona, the Brazilian Participation Group, Brookhaven National Laboratory, Carnegie Mellon University, University of Florida, the French Participation Group, the German Participation Group, Harvard University, the Instituto de Astrofisica de Canarias, the Michigan State/Notre Dame/JINA Participation Group, Johns Hopkins University, Lawrence Berkeley National Laboratory, Max Planck Institute for Astrophysics, Max Planck Institute for Extraterrestrial Physics, New Mexico State University, New York University, Ohio State University, Pennsylvania State University, University of Portsmouth, Princeton University, the Spanish Participation Group, University of Tokyo, University of Utah, Vanderbilt University, University of Virginia, University of Washington, and Yale University.

Funding for the Sloan Digital Sky Survey IV has been provided by the Alfred P. Sloan Foundation, the U.S. Department of Energy Office of Science, and the Participating Institutions. SDSS-IV acknowledges support and resources from the Center for High-Performance Computing at the University of Utah. The SDSS web site is \url{www.sdss.org}.

SDSS-IV is managed by the Astrophysical Research Consortium for the Participating Institutions of the SDSS Collaboration including the Brazilian Participation Group, the Carnegie Institution for Science, Carnegie Mellon University, the Chilean Participation Group, the French Participation Group, Harvard-Smithsonian Center for Astrophysics, Instituto de Astrof\'isica de Canarias, The Johns Hopkins University, Kavli Institute for the Physics and Mathematics of the Universe (IPMU) / University of Tokyo, Lawrence Berkeley National Laboratory, Leibniz Institut f\"ur Astrophysik Potsdam (AIP),  Max-Planck-Institut f\"ur Astronomie (MPIA Heidelberg), Max-Planck-Institut f\"ur Astrophysik (MPA Garching), Max-Planck-Institut f\"ur Extraterrestrische Physik (MPE), National Astronomical Observatories of China, New Mexico State University, New York University, University of Notre Dame, Observat\'ario Nacional / MCTI, The Ohio State University, Pennsylvania State University, Shanghai Astronomical Observatory, United Kingdom Participation Group, Universidad Nacional Aut\'onoma de M\'exico, University of Arizona, University of Colorado Boulder, University of Oxford, University of Portsmouth, University of Utah, University of Virginia, University of Washington, University of Wisconsin, Vanderbilt University, and Yale University.

We thank the E-Science and Supercomputing Group at Leibniz Institute for Astrophysics Potsdam (AIP) for their support with running the StarHorse code on AIP cluster resources.
\end{acknowledgements}


\bibliographystyle{aa}
\bibliography{apogee_galcen_dr16_revised}

\newpage

\appendix

\section{List of APOGEEIDs}
\begin{table}[!htbp]
  \caption{List of APOGEEIDs used in this paper for the analysis. The preliminary reduction version r13 has been used with the corresponding allStar-r13-l33-58672.fits file. This file  will be provided once the official data release DR17 becomes public.}
  \begin{tabular}{ccc}
    APOGEEID\\
    \hline
 2M17411696-2845379	& 2M17444724-2834457	& 2M17475364-2916499 \\
2M17415562-2854231	& 2M17445535-2822120	& 2M17475394-2933090 \\
2M17420041-2851435	& 2M17445978-2804160	& 2M17475847-2938176 \\
2M17420630-2846179	& 2M17450076-2831319	& 2M17480068-2939462 \\
2M17422063-2852202	& 2M17450464-2819092	& 2M17480297-2943223 \\
2M17422591-2901282	& 2M17450767-2840324	& 2M17480327-2923540 \\
2M17423476-2851278	& 2M17452445-2810402	& 2M17480578-2918285 \\
2M17423502-2849086	& 2M17452504-2808295	& 2M17480583-2925217 \\
2M17423812-2850540	& 2M17453070-2815270	& 2M17480772-2913302 \\
2M17423924-2840140	& 2M17453887-2833047	& 2M17480997-2924080 \\
2M17425347-2853162	& 2M17454263-2811016	& 2M17481444-2924176 \\
2M17430180-2837301	& 2M17454276-2805089	& 2M17481500-2923220 \\
2M17430420-2908362	& 2M17454586-2826088	& 2M17481744-2846454 \\
2M17430697-2904104	& 2M17455438-2819028	& 2M17481945-2828585 \\
2M17430939-2911452	& 2M17460129-2821004	& 2M17481975-2941196 \\
2M17431183-2805302	& 2M17460483-2948059	& 2M17482095-2931181 \\
2M17431215-2836211	& 2M17461894-2952180	& 2M17482165-2849161 \\
2M17431579-2837552	& 2M17462218-2831155	& 2M17482614-2903377 \\
2M17431676-2854588	& 2M17462609-2934081	& 2M17482665-2921284 \\
2M17432025-2835065	& 2M17462725-2935241	& 2M17482684-2939452 \\
2M17433138-2835128	& 2M17463316-2925543	& 2M17483299-2920129 \\
2M17433323-2825538	& 2M17463977-2915532	& 2M17483375-2936066 \\
2M17433389-2832086	& 2M17464346-2946229	& 2M17483681-2919095 \\
2M17433811-2833452	& 2M17465147-2936351	& 2M17483704-2926399 \\
2M17433895-2839578	& 2M17465235-2928073	& 2M17483843-2900451 \\
2M17433960-2803317	& 2M17470230-2945086	& 2M17483965-2907047 \\
2M17433967-2908204	& 2M17470331-2931348	& 2M17484023-2851445 \\
2M17434212-2810387	& 2M17470591-2925483	& 2M17484069-2836340 \\
2M17434430-2840387	& 2M17470766-2933217	& 2M17484277-2928243 \\
2M17434549-2819343	& 2M17471090-2927321	& 2M17484716-2902130 \\
2M17434740-2821036	& 2M17471201-2932176	& 2M17484806-2849357 \\
2M17434955-2835372	& 2M17471507-2928550	& 2M17484808-2919367 \\
2M17435231-2836460	& 2M17471743-2903217	& 2M17484937-2922212 \\
2M17435923-2828184	& 2M17471767-2927285	& 2M17485023-2857537 \\
2M17440216-2830017	& 2M17471773-2934473	& 2M17485464-2932428 \\
2M17440516-2825596	& 2M17471958-2901456	& 2M17485511-2908256 \\
2M17440701-2830436	& 2M17472118-2943123	& 2M17485976-2837160 \\
2M17441037-2827035	& 2M17472195-2932073	& 2M17490081-2906543 \\
2M17441210-2825556	& 2M17472783-2941543	& 2M17490234-2917537 \\
2M17441310-2829287	& 2M17473007-2939215	& 2M17490306-2912438 \\
2M17441358-2803383	& 2M17473184-2931553	& 2M17490612-2926138 \\
2M17441648-2844195	& 2M17473220-2942431	& 2M17490863-2922031 \\
2M17442423-2841005	& 2M17473329-2929159	& 2M17491129-2918514 \\
2M17442477-2830118	& 2M17473512-2944254	& 2M17491775-2902183 \\
2M17442705-2832044	& 2M17473827-2825007	& 2M17492165-2919289 \\
2M17442986-2838590	& 2M17473854-2931466	& 2M17492184-2909233 \\
2M17443100-2823357	& 2M17474071-2936281	& 2M17492271-2922456 \\
2M17443315-2830321	& 2M17474101-2925572	& 2M17492669-2909339 \\
2M17444047-2903581	& 2M17474135-2933431	& 2M17493231-2917440 \\
2M17444090-2914204	& 2M17474542-2934350	& 2M17493249-2916168 \\
2M17444213-2820294	& 2M17474997-2924254	& 2M17493309-2857103 \\
2M17444222-2841485	& 2M17475052-2926175	&    \\
2M17444388-2832377	& 2M17475074-2931190	&    \\
  \end{tabular}
  \end{table}

\end{document}